# Autothermal Reforming of Methane on Rhodium Catalysts: Microkinetic Analysis for Model Reduction

Marm Dixit[a], Renika Baruah[a], Dhrupad Parikh[a], Sudhanshu Sharma[a], Atul Bhargav[a,1]

[a]Energy Systems Laboratory, IIT Gandhinagar, Ahmedabad, Gujarat, INDIA

*Abstract*

Methane autothermal reforming has been studied using comprehensive, detailed microkinetic mechanisms, and a heirarchically reduced rate expression has been derived without apriori assumptions. The microkinetic mechanism is adapted from literature and has been validated with reported experimental results. Rate Determining Steps are elicited by reaction path analysis, partial equilibrium analysis and sensitivity analysis. Results show that methane activation occurs via dissociative adsorption to pyrolysis, while oxidation of the carbon occurs by O(s). Further, the mechanism is reduced through information obtained from the reaction path analysis, which is further substantiated by principal component analysis. A 33% reduction from the full microkinetic mechanism is obtained. One-step rate equation is further derived from the reduced microkinetic mechanism. The results show that the this rate equation accurately predicts conversions as well as outlet mole fraction for a wide range of inlet compositions.

[1] Corresponding author: Dr. Atul Bhargav, Assistant Professor (Mechanical Engineering)
A: IIT Gandhinagar (VGEC Campus), Visat-Gandhinagar Highway, Chandkheda, Ahmedabad, GJ 382424 INDIA
E: atul.bhargav@iitgn.ac.in | P: +91 814 030 7813

**Table 1:** Nomenclature

| Symbol | Description | Units |
|---|---|---|
| $\emptyset$ | partial equilibrium factor | - |
| $\theta_i$ | surface coverage | - |
| $C_i$ | species concentration | mol cm$^{-3}$ |
| $r_i$ | rate of reaction 'i' | mol cm$^{-3}$ s$^{-1}$ |
| S | sensitivity matrix | - |
| $S_{i,j}$ | sensitivity coefficient | - |
| $k_i$ | reaction constant | reaction specific |
| S/C | steam-to-carbon ratio | - |
| O/C | oxygen-to-carbon ratio | - |

**Acronyms**

| Symbol | Description |
|---|---|
| ATR | autothermal reforming |
| CFD | computational fluid dynamics |
| DFT | density functional theory |
| DR | dry reforming |
| PE | partial equilibrium |
| POx | preferential oxidation |
| RDS | rate determining step |
| RPA | reaction path analysis |
| SA | sensitivity analysis |
| SR | steam reforming |
| UBI-QEP | unity bond index-quadratic exponential potential |

# 1 Introduction

Hydrogen as an energy carrier[1], and fuel cells as energy conversion[2] devices have gained traction recently in the context of the shift toward more efficient and less carbon-intensive energy solutions [3–5]. This has renewed the interest in localized hydrogen generation with a focus on hydrocarbon reforming process [6–13], and methane is one of the prime sources of hydrocarbon based hydrogen [6,7,10–12,14]. Autothermal reforming as a means of hydrogen production has been gaining academic and research interest due to its thermodynamically neutral nature and feasible operating conditions [15–22].

In the past, much attention has been placed on the preparation of catalysts and the evaluation of the process and equipment with relatively little work being done on the numerical modelling, kinetics and mechanism of the reaction. Methane reforming is a system, where multiple reaction equilibriums are seen and hence several routes to the desired products and by products exist [23]. Thus, to optimize the reactor performance a fundamental understanding of the underlying surface catalytic phenomenon is required, which is provided by a detailed microkinetic mechanism [24]. A detailed microkinetic mechanism is further useful to accurately model the reaction over a wide range of reactor conditions [25]. A few groups such as Vlachos et al [26] and Deustchmann et al [27] have worked on the development of microkinetic mechanisms for steam reforming (SR), dry reforming (DR) and partial oxidation of Methane (POx) respectively using first principle methodologies such as UBI-QEP, DFT techniques and have validated these mechanisms with experimental findings.

While a microkinetic model enables quantification of the interaction between gas-phase and surface chemistry, it is also computationally expensive, especially when employed as part of CFD simulations, [28]. Hence it becomes essential to heirarchically reduce the detailed microkinetic mechanism to mechanisms that include only the major pathways of reforming. It can be further reduced to single-step or two-step rate equations that reduce the complex mathematical interpretation of the microkinetic models to simple algebraic equations [29–32]. Maestri et al. have a proposed a two-step overall rate expression for SR and DR of methane, by performing a top down hierarchical model reduction [26]. Partial oxidation of lean hydrocarbons were studied in detail by Vlachos et al. [25]. However, a top-down heirchical model reduction for methane autothermal reforming on rhodium catalyst is not seen in the literature.

In this paper, we have adopted a microkinetic mechanism from Deutschmann et al [27] and derive a single step rate equation for autothermal reforming of methane. Section two describes the analysis done on the full microkinetic model: comparision with experimental data, reaction path analysis (RPA), partial equilibrium analysis (PEA) and sensitivity analysis (SA) on the mechanism to determine the rate determining steps (RDS) and most abundant reaction intermediates (MARI). Section three details the proposed one-step overall rate expression for $CH_4$ ATR over Rh. We see good agreement of the reduced rate expression with the full microkinetic mechanism.

## 2 Analysis of the Full Microkinetic Mechanism

## 2.1 Comparision with Experimental Result

For an initial analysis, we adapted the microkinetic mechanism developed by Deutschmann et al for partial oxidation of methane on Rh catalyst comprising of 7 gas phase species, 12 surface species and 40 elementary like reactions [33]. The microkinetic mechanism employed is listed in Table 2. The thermo-kinetic calculator toolbox "Cantera [34]" (reactor module) was used to solve the plug flow model for autothermal reforming of methane on Rh catalyst. The conditions simulated mirrored those of experiments conducted by Ayabe et. al [35] (the reactant mixture consisted of 16.7% $CH_4$, 0–40.0% $H_2O$, 1.7–16.7% $O_2$, and balance $N_2$ (balance); the reactor dimensions and velocities were assumed to a space velocity of 7200 $h^{-1}$).

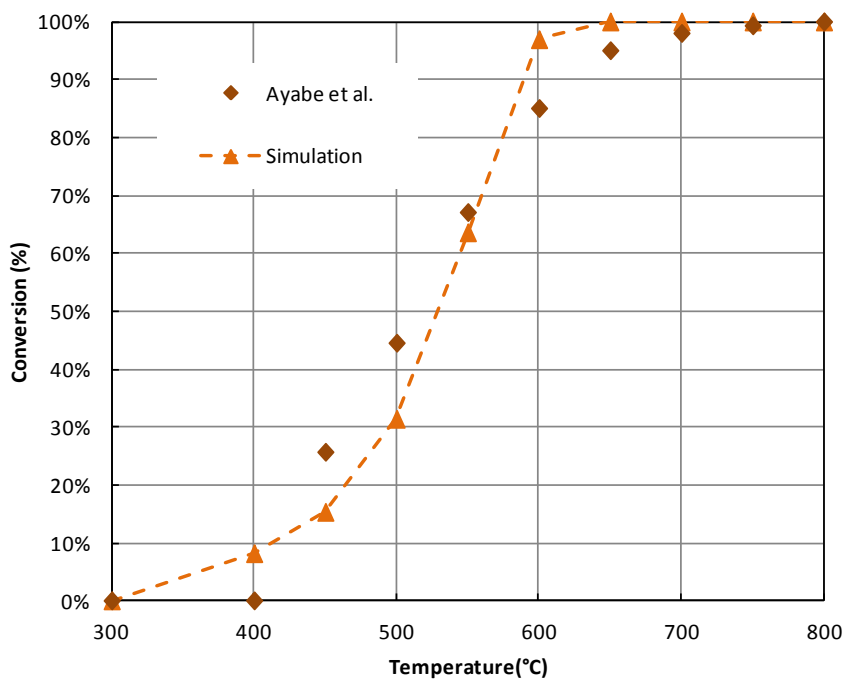

**Figure 1** Methane conversions with respect to temperature. The mole fractions of reactants are $CH_4:O_2:H_2O:N_2$ = 0.167:0.017:0.416:0.4.

The experiments were carried out in a fixed bed reactor with 2 wt% $Rh/Al_2O_3$ catalyst isothermally over a temperature range of 300°C to 800°C. Figure 1 shows a comparison of methane conversion over this temperature range between experimental and modelling results. It is evident that the results obtained from microkinetic model are in good agreement with experimental results.

**Table 2** Microkinetic Mechanism employed for the current study. It also lists shows the reactions included in the reduced mechanism

| Reaction Pair | No | Reaction $k_{rxn} = A \cdot exp\left(\frac{-E_a}{RT}\right)$ | Pre-exponential Factor or Sticking coefficient $A$ (s$^{-1}$) | Activation Energy $E_a$ (J/mol) | Reduced Mechanism |
|---|---|---|---|---|---|
| 1 | 1 | H2 + 2 Rh(s) -> 2 H(s) | 0.01 | - | ✓ |
|   | 2 | 2 H(s) -> H2 + 2 Rh(s) | 3.15E+21 | 77800 | ✓ |
| 2 | 3 | O2 + 2 Rh(s) -> 2 O(s) | 0.1 | - | ✓ |
|   | 4 | 2 O(s) -> O2 + 2 Rh(s) | 1.30E+22 | 35520 | ✓ |
| 3 | 5 | CH4 + Rh(s) -> CH4(s) | 0.008 | - | ✓ |
|   | 6 | CH4(s) -> CH4 + Rh(s) | 1.00E+13 | 25100 | ✓ |
| 4 | 7 | H2O + Rh(s) -> H2O(s) | 0.1 | - | ✓ |
|   | 8 | H2O(s) -> H2O + Rh(s) | 6.00E+13 | 45000 | ✓ |
| 5 | 9 | CO2 + Rh(s) -> CO2(s) | 1.0E-5 | - | ✓ |
|   | 10 | CO2(s) -> CO2 + Rh(s) | 3.00E+08 | 21700 | ✓ |
| 6 | 11 | CO + Rh(s) -> CO(s) | 0.5 | - | ✓ |
|   | 12 | CO(s) -> CO + Rh(s)             | 1.00E+13 | 133400 | ✓ |
|   | 13 | H(s) + O(s) -> OH(s) +Rh(s)     | 5.00E+22 | 83700  | ✓ |
| 7 | 14 | OH(s) +Rh(s) -> H(s) + O(s)     | 3.00E+20 | 37700  | ✓ |
|   | 15 | H(s) + OH(s) -> H2O(s) +Rh(s)   | 3.00E+20 | 33500  | ✓ |
| 8 | 16 | H2O(s) +Rh(s) -> H(s) + OH(s)   | 5.00E+22 | 104700 | ✓ |
|   | 17 | OH(s) +OH(s) -> H2O(s) + O(s)   | 3.00E+21 | 100800 | ✓ |
| 9 | 18 | H2O(s) + O(s) -> OH(s) +OH(s)   | 3.00E+21 | 171800 | ✓ |
|   | 19 | C(s) + O(s) -> CO(s) +Rh(s)     | 3.00E+22 | 97900  | ✓ |
| 10| 20 | CO(s) +Rh(s) -> C(s) + O(s)     | 2.50E+21 | 169000 | ✓ |
|   | 21 | CO(s) +O(s) -> CO2(s) + Rh(s)   | 1.40E+20 | 121600 | ✓ |
| 11| 22 | CO2(s) + Rh(s) -> CO(s) +O(s)   | 3.00E+21 | 115300 | ✓ |
|   | 23 | CH4(s) +Rh(s) -> CH3(s) + H(s)  | 3.70E+21 | 61000  | ✓ |
| 12| 24 | CH3(s) + H(s) -> CH4(s) +Rh(s)  | 3.70E+21 | 1000   | ✓ |
|   | 25 | CH3(s) +Rh(s) -> CH2(s) + H(s)  | 3.70E+24 | 103000 | ✓ |
| 13| 26 | CH2(s) + H(s) -> CH3(s) +Rh(s)  | 3.70E+21 | 44000  | - |
|   | 27 | CH2(s) +Rh(s) -> CH(s) + H(s)   | 3.70E+24 | 100000 | ✓ |
| 14| 28 | CH(s) + H(s) -> CH2(s) +Rh(s)   | 3.70E+21 | 68000  | - |
|   | 29 | CH(s) +Rh(s) -> C(s) + H(s)     | 3.70E+21 | 21000  | ✓ |
| 15| 30 | C(s) + H(s) -> CH(s) +Rh(s)     | 3.70E+21 | 172800 | - |
|   | 31 | CH4(s) + O(s) -> CH3(s) + OH(s) | 1.70E+24 | 80300  | ✓ |
| 16| 32 | CH3(s) + OH(s) -> CH4(s) + O(s) | 3.70E+21 | 24300  | ✓ |
|   | 33 | CH3(s) + O(s) ->CH2(s) + OH(s)  | 3.70E+24 | 120300 | - |
| 17| 34 | CH2(s) + OH(s) =>CH3(s) + O(s)  | 3.70E+21 | 15100  | - |
|   | 35 | CH2(s) + O(s) =>CH(s) + OH(s)   | 3.70E+24 | 114500 | - |
| 18| 36 | CH(s) + OH(s) =>CH2(s) + O(s)   | 3.70E+21 | 36800  | - |
|   | 37 | CH(s) + O(s) =>C(s) + OH(s)     | 3.70E+21 | 30100  | - |
| 19| 38 | C(s) + OH(s) =>CH(s) + O(s)     | 3.70E+21 | 136000 | - |
|   | 39 | CO(s) + H(s) =>C(s) + OH(s)     | 3.70E+21 | 143000 | - |
| 20| 40 | C(s) + OH(s) =>CO(s) + H(s)     | 3.70E+20 | 25500  | - |

## 2.2 Reaction Path Analysis and Rate Determining Step

A reaction path analysis helps identifying the main reaction paths involved in the process leading from reactants to products based on the net production rates of each surface species involved in the microkinetic mechanism[36]. The RPA was done with the aid of the object oriented programming tool within Cantera (and interfaced with python). We have run RPA for an isothermal reactor for cases where 0.5 < S/C < 2.5, 0.1 < O/C < 1.0 (both at inlet) and 600 < T < 800 °C, since moderate to high conversions were reported in the literature at these temperatures. [26].

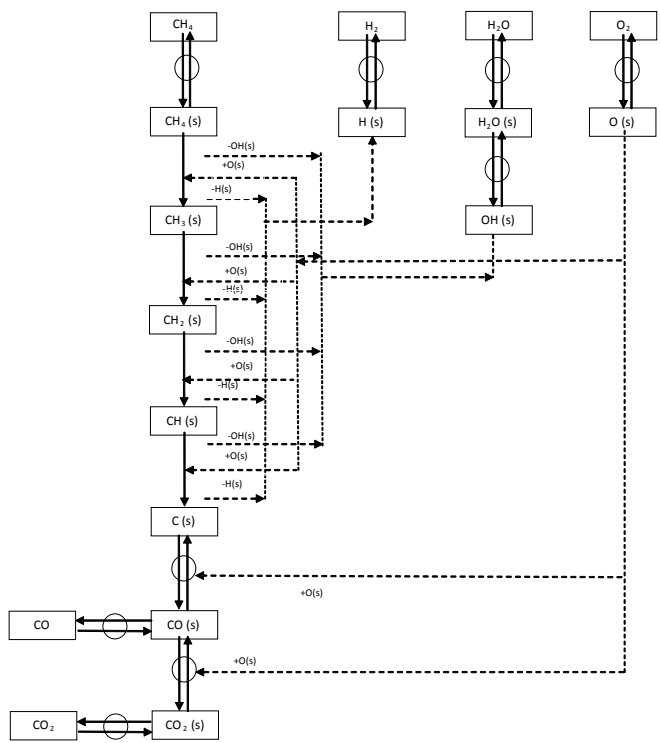

**Figure 2:** Reaction path diagram for S/C = 2.5, O/C = 0.1 at 600°C

RPA for S/C = 2.5 and O/C = 0.1 at 600 °C (Figure 2) shows that methane undergoes dissociative adsorption and dehydrogenates to C(s), which is expected. A number of experimental studies carried out on methane activation on rhodium surface have reported the activation of methane on rhodium surface via the dissociation route [37–40]. The C(S) is then further oxidised to CO(s) by O(s), which either gets desorbed, or is further oxidised to form $CO_2$(s), which eventually gets desorbed as $CO_2$. On increasing the O/C ratios, the complete oxidation pathways take over, with $CO_2$ and $H_2O$ being the predominant products. Increasing S/C ratios favors the partial oxidation product pathways. These trends have been seen experimentally for autothermal reforming in many studies [41,42]. Effect of temperature is seen prominently on the activation pathways of methane. At low temperatures, $CH_4$(s) is oxidised to $CH_3$(s) by OH(s) with successive dissociations happening pyrolytically. The oxidative dissociation by OH(s) if found to be important only for $CH_4$(s). For successive groups this pathway is diminished and completely eliminated at higher temperatures.

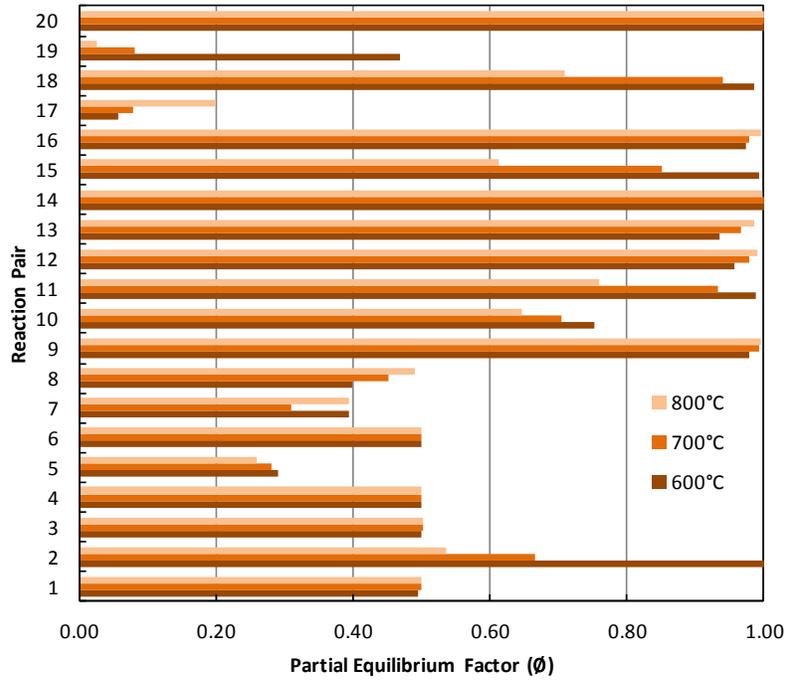

**Figure 3** Partial equilibrium analysis for S/C = 2.5 and O/C = 0.1 at different temperatures

Of the 40 reactions in the adopted microkinetic mechanism, the oxidative dissociation pathways were found to be negligible for all species, save for $CH_4(s)$. To further identify redundant reactions, and the rate determining step, a partial equilibrium analysis was carried out. We define partial equilibrium ratio as,

$$\phi = \frac{r_f}{r_f + r_b} \quad (1)$$

Where $r_f$ is the forward rate of reaction and $r_b$ is the backward rate of reaction. A value of ø=0.5 indicates that the reaction is in partial equilibrium while a deviation from 0.5 indicates reaction pairs not in equilibrium. Partial equilibrium analysis helps in identifying reactions that are the farthest from equilibrium, or the rate determining steps (RDS) based on Dumesic's criterion [43]. Figure 3 shows the PE analysis S/C = 2.5 and O/C = 0.1 at 600 °C. Similar results are observed for various S/C and O/C ratios as mentioned in section. Results indicate that almost all of the adsorption-desorption reactions are in partial equilibrium. However, a single RDS is difficult to be elicited from this.

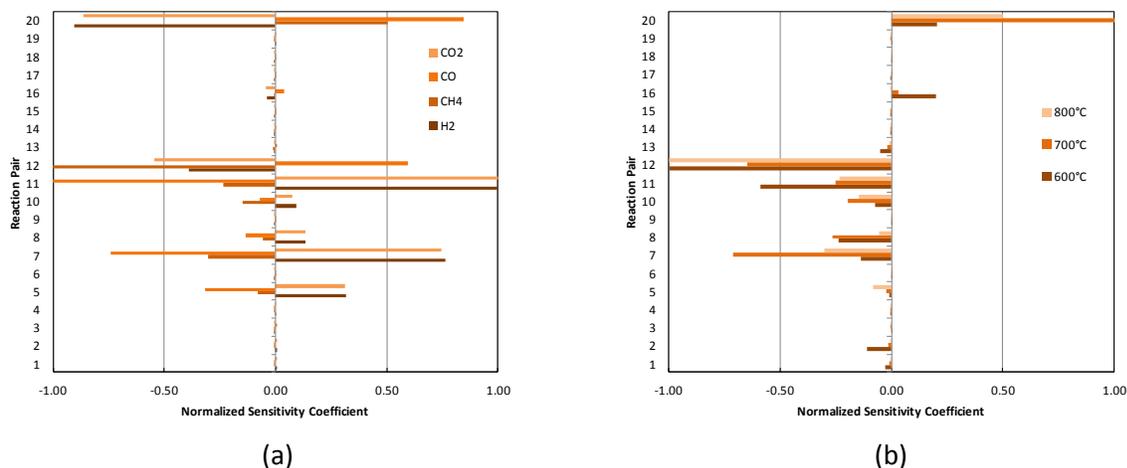

(a)            (b)

**Figure 4** Normalized sensitivity coefficients for S/C = 2.5 and O/C = 0.1. (a) Sensitivity of exit mole fractions at 600°C. (b) Sensitivity of exit mole fraction of $CH_4$ at different temperatures

Therefore, we have performed sensitivity analysis (SA) on outlet mole fraction of $CH_4$, $H_2$, CO and $CO_2$ by perturbing the reaction rates of each reaction pair by 5% (Figure 4). It is observed that reaction pair 11, 12 and 20 are most sensitive. Further sensitivity analysis of reactions for $CH_4$ outlet mole fraction between 600 and 800 $^o$C shows that the system is highly sensitive to reaction pair 12 (decomposition of $CH_4(S)$ to $CH_3(S)$ on Rh surface). Further on studying the PE results for various S/C and O/C cases, the $CH_4(s)$ dissociation to CH3(s) was found to be farthest from equilibrium. It has also been seen experimentally that methane activation is the sole kinetically important reaction [37]. Thus, the RDS for our mechanism is the reaction number 23, $CH_4(s) + Rh(s) => CH_3(s) + H(s)$.

Most Abundant Reactive Intermediate (MARI) is computed based on the surface coverage of each species on the surface of the catalyst. Figure shows the MARI for ATR of methane at S/C and O/C ratio of 2.5 and 0.1 respectively, at temperatures 600°C, 700°C and 800°C. CO(s) and C(s) are the MARI in all the three temperatures. As the temperature increases, a drop in surface coverages of CO(s) is observed while a slight increase in C(s) is recorded. With increase in S/C ratio, surface coverage of CO(s) rises and C(s) drops.

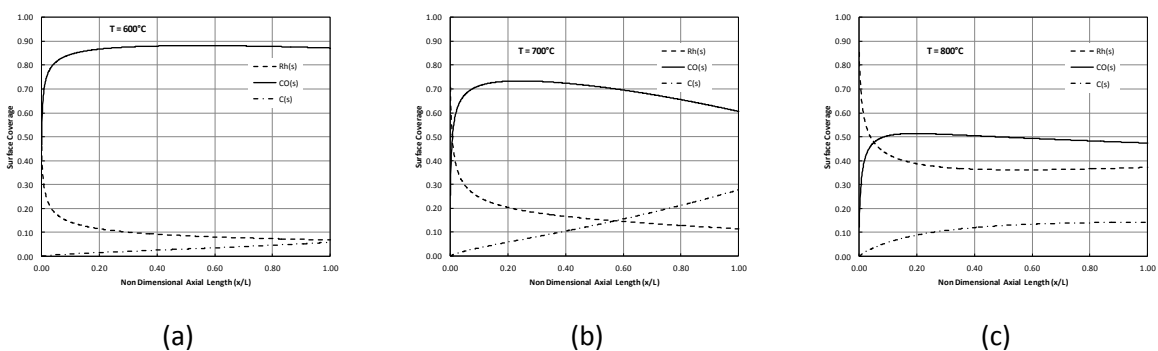

**Figure 5** Surface coverage along the reactor bed. The reactor conditions are S/C = 2.5 and O/C = 0.1. The surface coverages are recorded at 600°C, 700°C and 800°C

# 3 Model Reduction

In this section, we will reduce the full microkinetic model for the $CH_4/O_2/H_2O$ reacting system using the results from the RPA discussed earlier, while discarding all the elementary surface steps that play no apparently effective role. The reduction of microkinetic mechanism is further validated by performing a principal component analysis. With the reduced microkinetic mechanism in place, we derive a single step rate equation, by exploiting the information obtained from the reduced mechanism.

## 3.1 Reduced Microkinetic Mechanism

As was expected, not all steps detailed in the microkinetic mechanism are important: different pathways are activated with different reactor conditions. It is thus possible that only a select few of the full model can accurately model the ATR performance. The RPA indicates that with the exception of $CH_4(s)$ oxidation, the $CH_x(s)$ oxidation pathways are not important, with these pathways not being followed at all at higher temperatures. Furthermore, the partial equilibrium analysis suggests that all of the dehydrogenation reactions are far from equilibrium, with values of all reaction pairs above 0.9. This suggests that the forward reactions rates outweigh the reverse reaction rates, and hence the reverse reactions can be neglected from the full mechanism. Excluding the above reactions give us a 33% reduction from the full microkinetic mechanism.

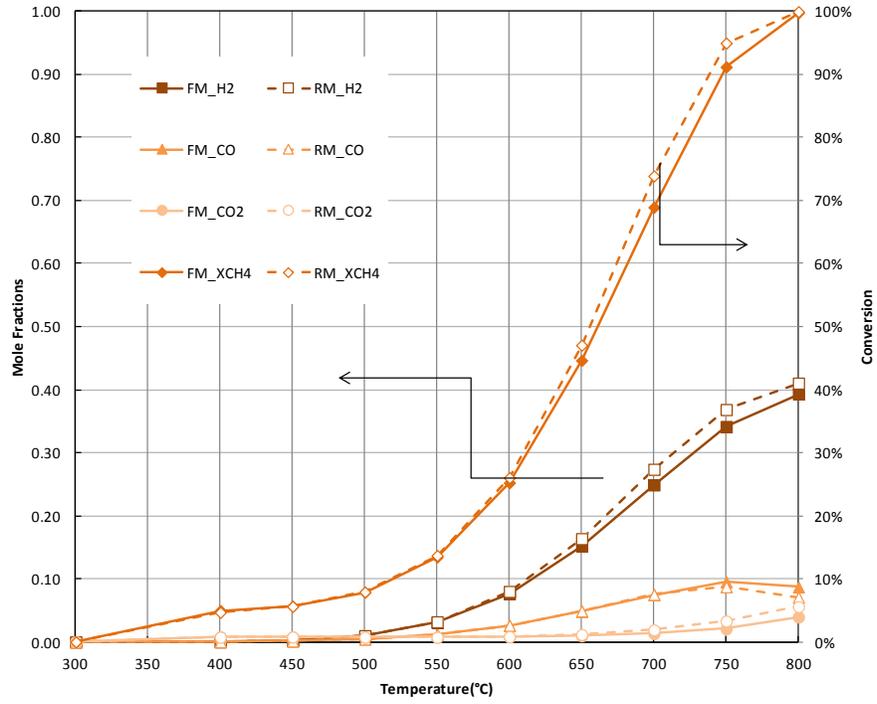

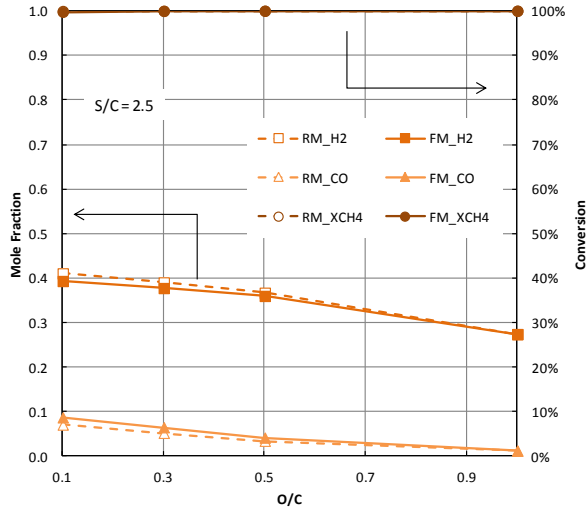
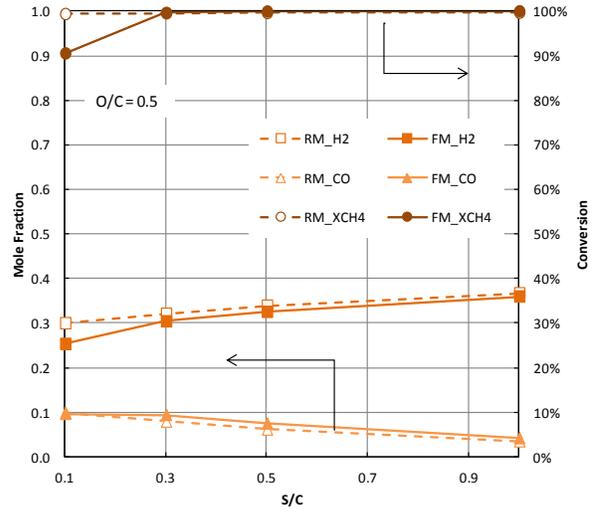

**Figure 6** (a) Temperature programmed reaction for S/C=2.5 and O/C = 0.1. Results show outlet mole fractions of CO, CO2 and H2 for full mechanism (FM) and reduced mechanism(RM). Methane conversions with temperature are also plotted. (b) Methane conversions, outlet H2, CO mole fraction for varying O/C for S/C= 2.5 (c) Methane conversions, outlet H2, CO mole fraction for varying S/C for O/C= 0.5.

Further, to substantiate this reduction, we performed principal component analysis using the information from the SA for multiple S/C and O/C ratios and three different temperatures [44,45]. We considered the matrix **S**, where individual elements $S_{i,j}$ represent the sensitivity coefficient of the outlet mole fractions of $i^{th}$ species to the $j^{th}$ reversible reaction. In particular, the mole fractions of $CH_4$, CO, $CO_2$, $H_2$ and $H_2O$ were monitiored for seven different cases as described earlier. Thus, a total of 105 sensitivities were computed for various temperatures and inlet concentrations. Since the full microkinetic mechanism consists of 20 reversible reaction pairs, the dimensions of the sensitivity matrix is 20 x 105, and the matrix S X $S^T$ is a 20 x 20 square matrix with 20 eigenvalues. Eigenvalues have been shown to provide an absolute measure of significant subsets of the full mechanism that consists of closely interacting elementary reactions. This information forms the basis of the objective criterion for selecting a minimal reaction set [45]. One of these eigenvalues is very important, and the corresponding eigen vector is then considered for choosing the important reaction pairs. The important reactions identified by the PCA and the RPA were found to be similar. The final reduced mechanism, which consisted of 27 elementary-like reactions are recorded in the last column of Table 2. The prediction of the full microkinetic model and the reduced microkinetic model are compared for various conditions in Figure 6. The results show excellent agreement for a wide range of operating conditions.

## 3.2 One step rate equation

Here, we use the reduced microkinetic model and attempt to derive an overall rate expression for ATR. The steady state balance equations for the surface species according to this model can be written as,

$$\frac{d\theta_H}{dt} = 2r_1 - 2r_2 - r_{13} + r_{14} - r_{15} + r_{16} + r_{23} - r_{24} + r_{25} - r_{26} + r_{27} - r_{28} + r_{29} - r_{30} \tag{1}$$

$$\frac{d\theta_{H2O}}{dt} = r_7 - r_8 + r_{15} - r_{16} + r_{17} - r_{18} \tag{2}$$

$$\frac{d\theta_{OH}}{dt} = r_{13} - r_{14} - r_{15} + r_{16} - 2r_{17} + 2r_{18} + r_{31} - r_{32} \tag{3}$$

$$\frac{d\theta_{CO}}{dt} = r_{11} - r_{12} + r_{19} - r_{20} - r_{21} + r_{22} \tag{4}$$

$$\frac{d\theta_{CO2}}{dt} = r_9 - r_{10} + r_{21} - r_{22} \tag{5}$$

$$\frac{d\theta_{CH4}}{dt} = r_5 - r_6 - r_{23} + r_{24} - r_{31} + r_{32} \tag{6}$$

$$\frac{d\theta_{CH3}}{dt} = r_{23} - r_{24} - r_{25} + r_{26} + r_{31} - r_{32} \tag{7}$$

$$\frac{d\theta_{CH2}}{dt} = r_{25} - r_{26} - r_{27} + r_{28} \tag{8}$$

$$\frac{d\theta_{CH}}{dt} = r_{27} - r_{28} - r_{29} + r_{30} \tag{9}$$

$$\frac{d\theta_{C}}{dt} = -r_{19} + r_{20} + r_{29} - r_{30} \tag{10}$$

$$\frac{d\theta_{O}}{dt} = 2r_{3} - 2r_{4} - r_{13} + r_{14} - r_{17} + r_{18} - r_{19} + r_{20} - r_{21} + r_{22} - r_{31} + r_{32} \tag{11}$$

$$\theta_{H} + \theta_{OH} + \theta_{H2O} + \theta_{CO} + \theta_{CO2} + \theta_{CH4} + \theta_{CH3} + \theta_{CH2} + \theta_{CH} + \theta_{C} + \theta_{O} + \theta_{RH} = 1 \tag{12}$$

Since the adsorption/desorption reactions are in partial equilibrium, the rates of forward and backward reactions are comparable. Thus, the coverages of the species H(s), O(s), CH4(s), H2O(s), CO2(s), CO(s) can be derived as follows:

$$\theta_{H} = \sqrt{\frac{k_{1}}{k_{2}} \cdot C_{H2}} \cdot \theta_{RH} \tag{13}$$

$$\theta_{O} = \sqrt{\frac{k_{3}}{k_{4}} \cdot C_{O2}} \cdot \theta_{RH} \tag{14}$$

$$\theta_{CH4} = \frac{k_{5}}{k_{6}} \cdot C_{CH4} \cdot \theta_{RH} \tag{15}$$

$$\theta_{H2O} = \frac{k_{7}}{k_{8}} \cdot C_{H2O} \cdot \theta_{RH} \tag{16}$$

$$\theta_{CO2} = \frac{k_{9}}{k_{10}} \cdot C_{CO2} \cdot \theta_{RH} \tag{17}$$

$$\theta_{CO} = \frac{k_{11}}{k_{12}} \cdot C_{CO} \cdot \theta_{RH} \tag{18}$$

Further, as previously stated, the steady state axial variation of surface coverages for different O/C and S/C ratios show that C(s) and CO(s) along with Rh(s) are the dominant surface species. Hence equation 12 can be simplified as

$$\theta_{CO} + \theta_{C} + \theta_{Rh} = 1 \tag{19}$$

Simplifying equation 10, we get the surface coverage of C(s) as follows,

$$\theta_C = \frac{K_{20} \cdot \theta_{CO} \cdot \theta_{RH} + K_{23} \cdot \theta_{CH4} \cdot \theta_{RH} + K_{31} \cdot \theta_{CH4} \cdot \theta_{RH}}{k_{10} \cdot \theta_{CO}} \qquad (20)$$

Hence, the surface coverage of Rh(s) can be given as,

$$\theta_{RH} = \frac{k_{19} \cdot \sqrt{\frac{k_3}{k_4} \cdot C_{O2}}}{k_{19} \cdot \sqrt{\frac{k_3}{k_4} \cdot C_{O2}} + \frac{k_{20} \cdot k_{11}}{k_{12}} \cdot C_{CO} + \frac{k_{23} \cdot k_5}{k_6} \cdot C_{CH4} + \frac{k_{31} \cdot k_5}{k_6} \cdot \sqrt{\frac{k_3}{k_4} \cdot C_{O2}} \cdot C_{CH4} + \frac{k_{19} \cdot k_{11}}{k_{12}} \cdot \sqrt{\frac{k_3}{k_4} \cdot C_{O2}} \cdot C_{CO}} \qquad (21)$$

Because the CH4 dehydrogenation reaction was found to be the RDS, the forward rate reaction for the ATR reaction is given by,

$$r_{atr,forward} = r_{23} = \frac{K_{23} \cdot K_5 \cdot}{k_6} \cdot \theta_{RH}^2 \cdot C_{CH4} \qquad (21)$$

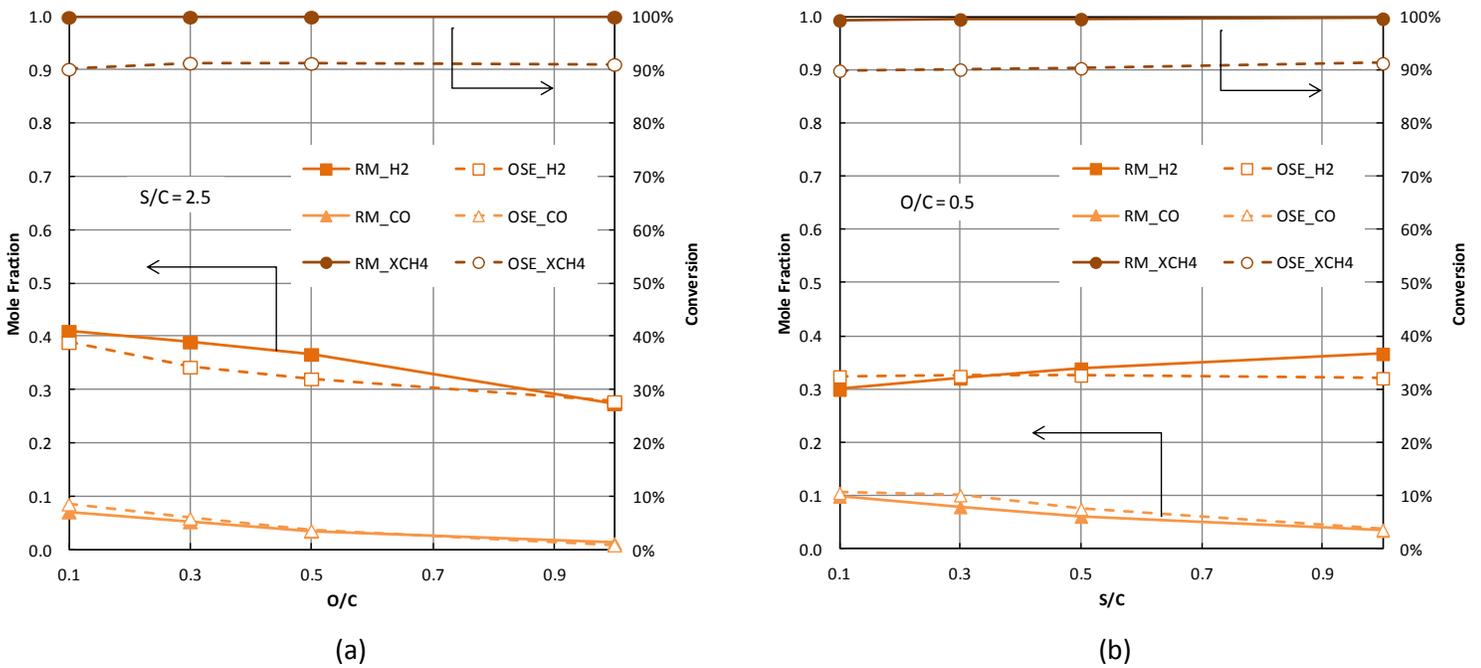

(a)     (b)

**Figure 7:** Comparision of reduced mechanism(RM) and one step rate equation (OSE). Methane conversion and the outlet mole fractions show good agreement with the reduced mechanism.

The reduced rate expression is compared with the full and reduced microkinetic mechanism for cases illustrated previously. It is seen that the reduced rate expression performs agreeably for a wide range of inlet concentrations and temperatures.

# 4 Conclusion

Methane ATR was analayzed using a validated microkinetic mechanism. Our analyses demonstrated the important reaction paths, the most abundant reaction intermediates and the key sensitive reaction steps. It is seen that regardless of the S/C and O/C ratios, the methane consumption ($CH_4 \rightarrow C(s) \rightarrow CO(s)$) is due to pyrolysis and carbon oxidation by $O(s)$. Further, in line with experimental results, methane activation is found to be sole kinetically relevant step. We have further derived a one-step rate equation based on the reduced microkinetic model. This equation is not based on prior assumptions, with parameters fitted to experimental data; rather, the rate is derived from the reduced microkinetic model itself. Our reduced mechanism is able to correctly predict the outlet mole fractions as well as the methane conversion for a wide range of S/C and O/C ratios reported in the literature.

# 5 Acknowledgments

The authors gratefully acknowledge the support received from IIT Gandhinagar, the Ministry of Human Resources, Government of India and the Ministry of Science & Technology (Grant number SR/S3/CE/078/2012 (DST)), Government of India.